\documentclass[aps,floatfix,twocolumn,amsmath]{revtex4}
\usepackage{graphicx}
\newcommand{\be}{\begin{equation}}
\newcommand{\ee}{\end{equation}}
\newcommand{\bea}{\begin{eqnarray}}
\newcommand{\eea}{\end{eqnarray}}

\begin{document}


\title{$f(R)$ Gravities \`a la Brans-Dicke}


\author{Tongu\c{c} Rador}
\email[]{tonguc.rador@boun.edu.tr}

\affiliation{Bo\~{g}azi\c{c}i University Department of Physics \\ 34342 Bebek, \.{I}stanbul, Turkey}


\date{\today}

\begin{abstract}
We extend $f(R)$ theories via the addition of a fundamental scalar field. The approach is reminiscent of the dilaton field of string theory and the Brans-Dicke model. $f(R)$ theories attracted much attention recently in view of their potential to explain the acceleration of the universe. Extending $f(R)$ models to theories with scalars can be motivated from the low energy effective action of string theory. There, a fundamental scalar  (the dilaton), has a non-minimal coupling to the Ricci scalar. Furthermore beyond tree level actions will contain terms having higher (or lower) powers of $R$ compared to the canonical Einstein-Hilbert term. Theories with $f(R)$ will contain an extra scalar degree on top of the ad-hoc dilaton and mixing of these two modes around a stable solution is a concern. In this work we show that no mixing condition mandates the form $V_{1}(\phi)f(R)+V_{2}(\phi)R^{2}$ for the action.
\end{abstract}


\maketitle

\section{Introduction}

Recent observations shows that the universe expands in an accelerated fashion. This can most easily be obtained by adding a cosmological constant (or vacuum energy). However unfortunately vacuum energy related to the present acceleration and the one predicted by theory differ by large orders of magnitude. Recently , theories of gravity which modify the Einstein-Hilbert action by adding terms with different powers of the Ricci scalar attracted much attention \cite{fr2} - \cite{zhuk5}. The motivation is that there is no ab-initio vacuum energy and one is free of the mentioned fine tuning problem.

Another modification of Einstein-Hilbert action is the Brans-Dicke model. In that approach there is a new fundamental field which couples non-minimally to the Ricci scalar and the coupling will yield Newton's constant along with the Einstein-Hilbert action if the field stabilizes at some value. This possible modification may seem arbitrary within the Brans-Dicke approach but string theory predicts a fundamental scalar called the dilaton and its role is very similar. In fact the two models can be related by a conformal transformation. So the motivation for this approach can be justified within a string theory inspired model. Furthermore it is conceivable that string theory will predict certain modification in the low energy limit of the theory. These modifications will generally be coming from beyond tree level calculations and involve higher powers of curvature quantities.

In this work we confine the study to 4 dimensions even though our conclusion is shown to trivially generalize to d-dimensions. We further consider models which contains  only algebraic functions of the Ricci scalar $R$. The motivation for this is that such models are free of the Ostrogradski \cite{os} (and \cite{os2} for a nice review) instability infesting theories with higher derivatives.

In a $f(R)$ theory there will be a scalar excitation. This excitation can be viewed to be the $\partial f/\partial R$ field. Adding another scalar on top of this brings in the issue of mixing if one expands around a solution with constant fields. As we will elaborate in the text, mixing is undesirable for a number of reasons so we confine this work to model with no mixing around a stable solution for the field equations. We show that the no-mixing condition strongly constrains the possible models. The general conclusion we advocate is that the most general model without mixing is the one without a pure potential term for the dilaton and $R$.

$f(R)$ theories with a an extra fundamental scalar field can be shown to be analogous to a bi-scalar tensor gravity theory via a conformal transformation of the original metric (see for example \cite{krep1} and \cite{krep2}). However, this transformation depends on the form of the function $f$ and constrains it a little bit. Furthermore in the conformally transformed equivalent tensor bi-scalar model the price for equivalency is an accompanying change in the original equations of motion for test bodies. This is expected since the test particles fundamentally feel the original metric. Such an approach based on a conformally transformed metric will bring in a theory with an equation of state for the scalars that depends on $f(R)$, as for instance shown in \cite{ref1} for pure $f(R)$ theories. There it was argued that data capable of selecting the true physical frame can
be used to discriminate such equivalent approaches. Due to these considerations we choose to work with the original metric ansatz without resorting to the conformally equivalent description.

\section{$f(R)$ Gravities with a scalar coupled non-minimally}

We choose the following action for this work,

\be
S=\int\;dx^{4}\;\sqrt{-g}\left[-\frac{1}{2}(\nabla \phi)^{2}+f(\phi,R)\right]\;.
\ee

The field equations resulting from this action can be cast as follows,

\begin{subequations}
\bea
\nabla^{2}\phi+f_{\phi}&=&0\\
\frac{1}{4}g_{\mu\nu}(\nabla\phi)^{2}-\frac{1}{2}\nabla_{\mu}\phi\nabla_{\nu}\phi +\tilde{G}_{\mu\nu}&=&0
\eea
\end{subequations}

\noindent Here subscript $\phi$ (or $R$) describe the partial derivative of $f$ with respect to $\phi$ (or $R$). We have defined the modified Einstein tensor as

\be
\tilde{G}_{\mu\nu}\equiv f_{R}R_{\mu\nu}-\frac{1}{2}fg_{\mu\nu}+\left(g_{\mu\nu}\nabla^{2}-\nabla_{\mu}\nabla_{\nu}\right)f_{R}
\ee

\noindent As usual the 0-0 part of the tensor equation can actually be traded for the trace equation 

\be
\frac{1}{2}(\nabla\phi)^{2}+f_{R}R-2 f+3\nabla^{2}f_{R}=0\;.
\ee

\subsection{Constant $\phi$ and $R$ solutions}

In this work we are interested in solutions with a constant $\phi=\phi^{o}$ and $R=R^{o}$. This can be achieved if one has

\begin{subequations}
\bea
f_{\phi}^{o}&=&0\;,\\
f_{R}^{o}R^{o}-2 f^{o}&=&0\;.
\eea
\end{subequations}

One also has to satisfy the tensor equation 

\be
f_{R}^{o}R_{\mu\nu}^{o}-g^{o}_{\mu\nu}\frac{1}{2}f^{o}=0\;\;.
\ee

\noindent Here and the above, the superscripts $o$ in $f$ and its derivatives mean that they are evaluated at $\phi^{o}$ and $R^{o}$.

In a cosmological scenario one usually takes the following for the metric (where we assumed that observed dimensions have no intrinsic curvature)

\be\label{eq:metrr}
ds^{2}=-dt^{2}+\sum_{i=1}^{3}\;e^{2A_{i}(t)}\;(dx^{i})^{2}\;\;.
\ee

\noindent In this ansatz the Ricci tensor is diagonal and the tensor equation written in the orthonormal frame for the solution we seek becomes 

\begin{subequations}
\bea
f_{R}^{o}R_{\hat{t}\hat{t}}^{o}+\frac{1}{2}f^{o}&=&0\\
f_{R}^{o}R_{\hat{i}\hat{i}}^{o}-\frac{1}{2}f^{o}&=&0
\eea
\end{subequations}

The meaning of this equation is clear. Unless $f^{o}=0$ and $f_{R}^{o}=0$ the space components of the Ricci scalar must be "numerically" the same if they are in general allowed to be different. It is rather interesting because unlike Einstein gravity this seem to imply non-trivial {\bf isotropization}. Thus the solution we are interested in becomes $R_{\mu\nu}^{o}=g_{\mu\nu}^{o}R^{o}/4$. From now on we are taking $A_{i}(t)=A(t)$.

The solution $R_{\mu\nu}^{o}=g_{\mu\nu}^{o}R^{o}/4$ clearly does not depend on the explicit form of the metric ansatz and is quite general in $f(R)$ theories. For instance one of the motivations to study $f(R)$ models is the existance of de Sitter or anti-de Sitter black holes with arbitrary horizon topology \cite{ref2}-\cite{ref3}. Such solutions must exist even with our  inclusion an extra scalar to the theory. However, since our main motivation
in this work is the FRW metric ansatz (\ref{eq:metrr}) we do not pursue this avenue further.

\subsection{Stability of the solution}

If one can find a solution satisfying the conditions one has to study whether small perturbations around the solution are not growing. That is the solution is stable, at least in the linear sense. There are two aspects to the study of stability. First the metric scale factors $A_{i}(t)$ must not grow under arbitrary perturbations around the solution $A_{i}^{o}(t)$. Second the two scalars that are in the theory must not have negative mass squared. One of these scalars is of course $\phi$, the other is the scalar that is
activated in $f(R)$ theories, namely $f_{R}$.

Under small perturbations we have 

\[
\nabla^{2}\delta\phi=-f^{o}_{\phi\phi}\delta\phi-f^{o}_{R\phi}\delta R\;,
\]

\noindent and

\[
3f^{o}_{RR}\nabla^{2}\delta R + 3f^{o}_{R\phi}\nabla^{2}\delta\phi=(f^{o}_{R}-f^{o}_{RR}R^{o})\delta R-f^{o}_{R\phi}\delta\phi\;.
\]

This can be cast into the following form 

\be
\nabla^{2}\delta X=\frac{1}{3 f^{o}_{RR}}\;S\delta X
\ee

\noindent with $\delta X^{T}=(\delta R,\delta\phi)$ and the elements of stabilization matrix $S$ is given by

\begin{subequations}
\bea
S^{RR}&=&f^{o}_{R}-f^{o}_{RR}R^{o}+3 {f^{o}_{R\phi}}^{2}\\
S^{\phi\phi}&=&-3 f^{o}_{\phi\phi}f^{o}_{RR}\\
S^{R\phi}&=&f^{o}_{R\phi}\left[-R^{o}+3f^{o}_{\phi\phi}\right]\\
S^{\phi R}&=&-3 f^{o}_{\phi R}f^{o}_{RR}
\eea
\end{subequations}

For the stability of the $\phi$ and $f_{R}$ excitations the matrix $S$ must have positive singular values regarding the fact that it is in general not symmetric.

Apart from requiring positive eigenvalues for $S$ there is also the issue of mixing. The field $\phi$ is an analogue of the dilaton field predicted by string theory \footnote{Even though we did not take the canonical form of the low energy string action our field $\phi$ can be considered to be directly related to the dilaton of string theory by a conformal transformation which is well behaved generally in contrast to the possibility of not being able to perform the conformal transformation for every $f$ in a higher order gravity theory}. Dilaton stability is
an important aspect of string theory motivated models. For example a rolling dilaton would imply a time-dependent Newton's constant which is subject to stringent bounds. These bounds can be met within a specific model, however for purely aesthetical reasons we would like to seek a possibility where
it never occurs.
On the other hand if there is a mixing between the dilaton and $f_{R}$ around a constant $\phi$ solution this would imply that  taking the dilaton to be a constant at the action level and assuming that the resulting theory is a stable theory would be contradicting arguments. Our main emphasis in this discussion is that the field $\phi$ is supposed to be a fundamental scalar field as opposed to the fact that $f_{R}$ excitations are genuinely gravitational in origin. We would like to stress this point further by invoking the standard model of particle physics. Gravity is assumed to be a singlet under the electro-weak symmetry group and this fact is also consistent with the anomaly cancellations. However if one would like to contemplate the possibility that $\phi$ is not a singlet under the electro-weak symmetries, a mixing between $\phi$ and $f_{R}$ can be very problematic.

So it is somewhat desirable to prevent mixing.
As it is, there seems to be no natural mechanism \footnote{That is without contrived cancellations between the quantities $f^{o}$, $f^{o}_{R}$ etc.} to prevent $\phi$ and $f_{R}$ mixing unless $f^{o}_{R\phi}=f^{o}_{\phi R}=0$. This can be trivially achieved if there are no non-minimal couplings between $\phi$ and $R$ but this will immediately take us outside the emphasis of this work. Another instance of no mixing can be provided if the extremum values $\phi_{o}$ and $R_{o}$ are not related by the conditions but this can trivialize the model. We will see in the next sections that there is a way to have no mixing without fine tuning.

Another important point is that the excitations $\delta\phi$ and $\delta R$ must not have vanishing masses as this is strongly disfavored from a phenomenological point of view.

There are various terms that play a privileged role in finding solutions, providing stability and preventing mixing. Terms that depend only on $R$ can have no bearing on the extremum condition in the $\phi$ equations. Pure $\phi$ terms on the other hand would be important for both sectors. Pure $\phi$ and pure $R$ terms cannot directly influence the easiest no-mixing condition $f^{o}_{R\phi}=0$. Furthermore the only term that would not affect the extremum condition of the trace equation and the $\phi$ equation is the ({\em globally}) conformally invariant $R^{2}$ term. This term will only influence the stability condition. As such it is an important term to possibly save a model without stability as the role of the $R^{2}$ term to save $R+1/R$ gravity models.

\section{Models Involving $R^{p}$ with $p<1$ }

In this section we consider several models of interest involving powers of $R$ such that they are more important near $R=0$. These models attracted much attention recently in view of their potential of explaining the present acceleration of the universe.

\subsection{Models That Won't Work}
As a first try we study the following

\be
f(\phi,R)=\alpha \phi^{2n1+2} R-\beta \phi^{2n2+2} R^{-m}+\gamma R^{2}
\ee

Since the $R$ term is not present on its own this is a flavor of the Brans-Dicke approach. We have also tried to keep the $\phi\to-\phi$ symmetry of the kinetic term of the scalar field. It is special in the sense that it does not have a pure $\phi$ potential and hence no analogue of a vacuum energy (cosmological constant). This is aesthetically favorable since the main purpose of $f(R)$ theories is to explain the acceleration without a cosmological constant. As we have mentioned $R^{2}$ term has no role in finding an extremum and can possibly save the model if without it there is instability. The extremum conditions are easily met if 

\begin{subequations}
\bea
R^{m+1}\phi^{2n_{1}-2n_{2}}&=&(2+m)\frac{\beta}{\alpha}\;,\\
(n_{2}+1)&=&(m+2)(n_{1}+1)\;.
\eea
\end{subequations}

\noindent We immediately see one important feature. The extremum condition for the trace equation is trying to conformalize the extremum and bringing a constraint on $n_{2}$ and $n_{1}$. By conformalization we only want to make a reference to the fact that $R^{2}$ term cannot influence the extremum condition of the trace equation and that it is conformally invariant. The extremum conditions are doing their best to resemble the rest of the action to that form. Of course such an effect cannot be present in a theory where there is no $\phi$ field. Since $n_{2}$ and $n_{1}$ are related this has direct bearing on the dimensionalities of the coupling constants. In fact one can show that for all $m$ one has ${\rm det}(S)=0$ \footnote{This is also clear from the fact that the extremum conditions do not fix $R^{o}$ and $\phi^{o}$ separately} meaning there will be a massless excitation which is not phenomenologically favored. This problem has occurred because there was no independent mass scale in the theory. Since there are two conditions for extremum one of them served to relate the dimensionalities of $\alpha$ and $\beta$. 

We therefor have to add terms which introduces two ad-hoc mass scales which will not be constrained by the extremum conditions. To exemplify the need for this let us try the following where there is only one possible mass scale,

\be
f(\phi,R)=\alpha\phi^{2} R +\lambda \phi^{4} -\beta \phi^{2n+2}R^{-m}+\gamma R^{2}\;.
\ee

The extremum conditions give

\begin{subequations}
\bea
-\alpha R - 2\lambda\phi^{2}+(m+2)\beta\phi^{2n}R^{-m}&=&0\;,\\
\alpha R +2\lambda\phi^{2}-(n+1)\beta\phi^{2n}R^{-m}&=&0\;. 
\eea
\end{subequations}

\noindent which implies 

\[
m+2=n+1
\]

\noindent and again $\phi_{o}$ and $R_{o}$ cannot be fixed separately.

\subsection{A Possible Model}

It is now apparent that we have to introduce two mass scales that are not related via the extremum conditions. If one adds a mass term for the scalar field and the usual Einstein-Hilbert term this will be trivially done. But we would be introducing an analogue of a vacuum energy and also go outside the Brans-Dicke approach. One way out of this dilemma that might generally work is to never allow pure potentials for the $\phi$ field, that is every term in the action will have to be an interaction between $\phi$ and $R$. For example the following 

\be
f(\phi,R)=\alpha \phi^{2} R -\beta \phi^{2n+2}\frac{1}{R}+\omega\phi^{2p+2}R^{2}+\gamma R^{2}\;.
\ee

could work because one fixes 

\begin{subequations}
\bea
R_{o}&=&\frac{(n-2)\alpha}{3\omega (p+1)}\;\phi_{o}^{-2p}\;,\\
\phi_{o}^{2n+4p}&=&\frac{(n-2)^{2}\alpha^{3}}{27\beta\omega^{2}(p+1)^{2}}.
\eea
\end{subequations}

\noindent as usual $\gamma$ term can be used to save stability. We would like to  focus on this model a little more. Taking $\alpha>0$ is important because this term is related to the effective Newton's constant. On the other hand, since we would like to have $R_{o}>0$ to account for the acceleration of the universe and that $2n+4p$ is even we get the following bounds on the parameters

\begin{subequations}
\bea
\beta&>&0\;,\\
\frac{(n-2)}{\omega (p+1)}&>&0\;.
\eea
\end{subequations}

Using the extremum values one can check the mixing status of the model. It is rather interesting that 

\[
{\rm No\;\;\;Mixing\;\;\;}\Longleftrightarrow\;n=0\;\;.
\]

That is the term with the $1/R$ will have to be a dimension zero operator, the most relevant term possible in terms of dimensional analysis. Dimension zero operators can only occur when there are more than one field around and possible inverse powers are in the game. Since this is not a fine tuning on the coupling constants it is legitimate to assume it. This will imply

\begin{subequations}
\bea
\beta&>&0\;,\\
\omega (p+1)&<&0\;.
\eea
\end{subequations}

The absence of mixing makes the stability analysis much simpler since $S$ is diagonal from which we can read the masses as follows,

\begin{subequations}
\bea
{m_{f_{R}}}^{2}=\frac{1}{3} \left[\frac{f^{o}_{R}}{f^{o}_{RR}}-R^{o}\right]\;\;. \\
{m_{\phi}}^{2}=-\frac{1}{3} f^{o}_{\phi\phi}\;\;.
\eea
\end{subequations}

The general condition for positivity of the mass of the $\phi$ field yields $p>0$. The other condition requests that $\gamma>\gamma_{o}$ where $\gamma_{o}$ depends on the parameters of the system. Since $p>0$ the positivity of $R_{o}$ imposes $\omega<0$. This is only fair because it is  the term with the  largest power of $\phi$ in the model and if we consider the potential for a given value of $R$ it must be bounded. 

To conclude the study of this model we cast it into a
standard form where every parameter is positive and the dimensionalities are apparent

\[
f(\phi,R)=a\phi^{2}R-\mu_{1}^{4}\frac{\phi^{2}}{R} - \left(\frac{\phi}{\mu_{2}}\right)^{2m+2}R^{2}+c R^{2}
\]

\noindent Here all the couplings $a$, $b$, $w$ and $c$ are dimensionless. The dimensions are described by $\mu_{1}$ and $\mu_{2}$. Let us further investigate the behaviour of the model for the particular instance of $m=1$. In this case we have

\begin{subequations}{\label{eq:modelll}}
\bea
R_{o} &=& \sqrt{\frac{3}{a}} \mu_{1}^{2} \;,\\
\phi_{o} &=& \mu_{2}\sqrt{s}\;,\\
m_{\phi}^{2} &=& 8\sqrt{\frac{a}{3}}\mu_{1}^{2}\;,\\
m_{f_{R}}^{2} &=& \frac{a}{3}\mu_{2}^{2}\frac{s}{c-2s^{2}}\;,\\
s&\equiv&\left(\frac{\mu_{2}}{\mu_{1}}\right)^{2}\sqrt{\frac{a^{3}}{27}}\;.
\eea
\end{subequations}

If one fixed the Hubble constant $H$ from experiments one fixes $R_{o}$ and hence $\mu_{1}$, this number is rather small. It follows immediately that one of the excitations will have to have a mass of the same scale as the Hubble constant. Such a particle if it exists must couple very weakly to matter. In fact this somewhat mandates to assume that it is a singlet under electroweak and strong interactions and therefor couples to the ordinary fields only via gravity. This possible handicap actually exists in all $f(R)$ theories with or without a Brans-Dicke type approach. However in a pure $f(R)$ approach one is not forced to assume the excited field $f_{R}$ is a singlet under standart model interactions: it is already a singlet from the start. So the general conclusion is that the Brans-Dicke type field must be a singlet under standard model unless rather contrived and ad-hoc hierarchies are imposed. Of course in this discussion we have assumed that the dimensionless coupling $a$ is of the order unity we would like to emphasize that if $a$ is allowed to vary from unity, as apparent  from (\ref{eq:modelll}), $R_{o}$ and $m_{\phi}^{2}$ can be very different from each other. This of course will bring in the issue of naturalness similar to the $\omega$ parameter of the standard Brans-Dicke model.

\subsection{Most General Model Without Mixing}

In an attempt to generalize the model presented above one can try the following

\be
f(\phi,R)=\alpha\phi^{2}R-\beta\phi^{2n+2}R^{q}-\omega\phi^{2p+2}R^{2}+\gamma R^{2}\;\;.
\ee

\noindent with $q<1$. A rather interesting feature is that if one searches for a no-mixing condition we get the same result,

\[
{\rm No\;\;\;Mixing\;\;\;}\Longleftrightarrow\;n=0\;\;\;\;{{\rm for\;\;\;any\;}\;\;\;q}
\]

The rest of the analysis is similar. The requirement that $R_{o}>0$ will mean $\omega>0$. The stability conditions will fix $p>0$ and $\gamma>\gamma_{o}$ where $\gamma_{o}$ is model dependent.

So if one has a working pure $\tilde{f}(R)$ mode, the quest to embed it in a Brans-Dicke approach must start with the following

\be
f(\phi,R)=\phi^{2}\tilde{f}(R)-\omega \phi^{2p+2}R^{2}+\gamma R^{2}\;\;.
\ee

The reason why the method exposed works is of course due to the fact that we have used the conformally invariant term $R^{2}$ which will not enter the extremum equations on $R$. So a working $\tilde{f}(R)$ gets us half the way without any problem. In the absence of $V$ the condition of no-mixing and the extremum equation on $\phi$ are identical so we get two birds with one stone. 

In fact the most general model without mixing is the following

\be
f(\phi,R)=V_{1}(\phi)\tilde{f}(R)+V_{2}(\phi)R^{2}
\ee

\section{Conclusion}

In this work we have studied a way to get $f(R)$ gravities via a scalar field which could be motivated by the dilaton of string theory.
We have made emphasis on avoiding the mixing between the fundamental scalar and the scalar $f_{R}$ that is excited in an $f(R)$ theory. One could of course relax this requirement and still can find a viable model. However we have shown that the no mixing requirement yielding the most general model actually kills two birds with one stone. That is the extremum condition on the fundamental scalar of the model and the no mixing condition become identical and hence there is no fine tuning of parameters. 

Even though we have confined our study to 4-dimensional models generalization of the no-mixing model can be easily done. The following will be free of mixing in d-dimensions.

\be
f(\phi,R)=V_{1}(\phi)\tilde{f}(R)+V_{2}(\phi)R^{d/2}
\ee

We have shied away from adding a pure $\phi$ potential on grounds that it would complicate the models beyond the scope of this work. Furthermore adding such a term also opens the question about a pure $R$ term (other than $R^{2}$) and this will take us outside the Brans-Dicke approach. Nevertheless it is an important extension of the models presented here.

In this work we have worked with a usual kinetic term for the $\phi$ field. It will be interesting to use the canonical form of the dilaton action coming from string theory and extend the study there. The motivation for this will be to provide a stabilization mechanism for the dilaton. So far in the literature the majority of approaches are focusing on dealing with the stability of the dilaton via energy-momentum tensors of compact objects around extra dimensions. It could be interesting to study the impact of a $f(R)$ approach in string inspired dilaton theory as it could yield non-canonical ways to stabilize the dilaton.

\end{document}